\documentclass[pre,twocolumn,a4paper,superscriptaddress,groupedaddress,floatfix]{revtex4}

\usepackage{graphicx}
\usepackage{dcolumn}
\usepackage{bm}
\usepackage{color}
\usepackage[mathlines]{lineno}
\usepackage{rotating}
\usepackage[]{amsmath}
\newcommand{\B}[1]{{\bm{#1}}}
\def\<{\langle}
\def\>{\rangle}
\def\tg{\tau_{\rm g}}
\def\trd{\tau_{\rm rd}}
\def\tsd{\tau_{\rm sd}}
\def\tmf{t_{\rm mf}}
\def\vcg{v_{\rm cg}}

\def\Pe{{\rm Pe}}
\def\t{t_{\rm coll}}

\begin{document}

\title{Stability phase diagram of active Brownian particles}

\author{Pin Nie}
\affiliation{ School of Physical and Mathematical Science, Nanyang Technological University, Singapore}
\affiliation{Singapore-MIT Alliance for Research and Technology, Singapore}
\author{Joyjit Chattoraj}
\affiliation{ School of Physical and Mathematical Science, Nanyang Technological University, Singapore}
\affiliation{ Institute of High Performance Computing, Agency for Science, Technology and Research,  Singapore}
\author{Antonio Piscitelli}
\affiliation{ School of Physical and Mathematical Science, Nanyang Technological University, Singapore}
\affiliation{CNR--SPIN, Dipartimento di Scienze Fisiche,
Universit\`a di Napoli Federico II, I-80126, Napoli, Italy}
\author{Patrick Doyle}
\affiliation{Singapore-MIT Alliance for Research and Technology, Singapore}
\affiliation{Department of Chemical Engineering, Massachusetts Institute of Technology, Cambridge, Massachusetts,
USA}
\author{Ran Ni}
\email{r.ni@ntu.edu.sg}\emph{}
\affiliation{School of Chemical and Biomedical Engineering, Nanyang Technological University}
\author{Massimo Pica Ciamarra}
\email{massimo@ntu.edu.sg}
\affiliation{ School of Physical and Mathematical Science, Nanyang Technological University, Singapore}
\affiliation{CNR--SPIN, Dipartimento di Scienze Fisiche,
Universit\`a di Napoli Federico II, I-80126, Napoli, Italy}

\date{\today}

\begin{abstract}
Phase separation in a low-density gas-like phase and a high-density liquid-like one is a common trait of biological and synthetic self-propelling particles' systems. 
The competition between motility and stochastic forces is assumed to fix the boundary between the homogeneous and the phase-separated phase.
Here we demonstrate that, on the contrary, motility does also promote the homogeneous phase allowing particles to resolve their collisions.
This new understanding allows quantitatively predicting the spinodal-line of hard self-propelling Brownian particles, the prototypical model exhibiting a motility induced phase separation.
Furthermore, we demonstrate that frictional forces control the physical process by which motility promotes the homogeneous phase. Hence, friction emerges as an experimentally variable parameter to control the motility induced phase diagram.
\end{abstract}

\maketitle

\section{Introduction}
Many biological and synthetic systems of self-propelled particles exhibit a transition from a homogeneous state to
one in which a gas- and a liquid-like phase coexist.
~\cite{Fily2012, Marchetti2013, Bechinger2016}.
While diverse physical processes might be responsible for the observed transition, the bare presence of motility is enough to induce it~\cite{Cates2015}.
Indeed, motility induced phase separation (MIPS) occur in systems of particles whose interactions are purely repulsive and do not promote the alignments of the self-propelling directions.
The prototypical simulation model is the active Brownian particles(ABP) model, which consists of spherical self-propelled particles interacting via excluded volume forces, and subject to thermal noise~\cite{TenHagen2011, Romanczuk2012, Fily2012, Redner2013, Speck2014, Speck2016}.

In active systems, two particles colliding head-to-head, or nearly so, severely slow-down their motion, reducing the local pressure.
This pressure drop may seed a positive-feedback mechanism leading to the formation of a dense cluster of active particles, and hence to phase separation~\cite{Takatori2015,Solon2015}. 
A similar scenario occurs in granular systems, where the pressure drop is due to the dissipative nature of the interparticle collisions~\cite{McNamara1992,Goldhirsch1993}. A homogeneous system of active particles is not always unstable towards phase separation, as there are physical processes that promote the homogeneous phase, opposing the above instability mechanism. The balance between the mechanisms promoting phase separation, and those promoting the homogeneous phase, sets the limit of stability of the homogeneous phase in the motility-density plane.

The rotational diffusion plays a role because, before a collision seed the growth of a cluster, the two colliding particles may change their self-propelling direction, and swim away~\cite{Redner2013, Buttinoni2013, Wysocki2014, Cates2015, Redner2016}.
This process gives rise to a flux of particles from the dense to the less dense phase promoting the homogeneous phase.
By balancing this flux and the reverse flux of particles migrating towards the denser phase, which is controlled by the activity, Redner {\it et al.}~\cite{Redner2013, Redner2016} predicted a low-density coexistence line of ABPs in good agreement with numerical results but did not predict the location of the critical point and the upper coexistence line. 
The other possibility is that the translational rather than the rotational noise promotes the homogeneous phase.
This scenario is suggested by a continuum equation for the evolution of the coarse-grained density and polarization fields~\cite{Fily2012, Zottl2013, Bialke2013, Speck2014, Fily2014, Stenhammar2014},
which is formally related to a thermodynamic approach aiming to map active Brownian particles into a equilibrium system~\cite{Solon2015}. 
This scenario predicts U-shaped spinodal line in the activity-density plane resulting from a diffusive instability. 
The prediction correctly reproduces the divergence of the lower spinodal line at a finite density, as well as the existence of a critical point.
However, this approach underestimates~\cite{Cates2015} the minimum value of the activity at the critical point by a factor $\simeq 10$. The limitations~\cite{Cates2015} of theoretical approaches based on the rotational rather than on the translational diffusivity in predicting the phase diagram of APBs, suggests that additional processes promoting phase separation may exist.

In this manuscript, we demonstrate a physical process promoting the homogeneous phase driven by the motility of the particles. Motility, therefore, promotes and opposes phase separation at the same time.  
We formalize this and the other mechanisms promoting and opposing phase separation in a collisional framework and predict the spinodal line of ABPs. 
Our prediction favourably compares to both two- and three-dimensional numerical simulations, for different values of the control parameters.
Furthermore, we demonstrate that friction tunes the features of the motility induced phase diagram, as it controls the new instability mechanisms we have uncovered.

\section{Kinetic model}
We develop a kinetic model to predict the spinodal line of ABPs particles. In this model, the dynamics is described by the following overdamped equation of motion:
\begin{eqnarray}
{\B {v}}_{i} &=& \frac{{\B F}_i}{\gamma}+ \frac{F_a}{\gamma} {\B n}_{i} + \sqrt{2D_t}\B\eta_i^t \label{eq:newton_t} \\
\dot {\B n}_{i} & =&  \sqrt{2 s D_r}\B\eta_i^r \times {\B n}_{i}.
\label{eq:newton_r}
\end{eqnarray}
Here $s D_r$ and $D_t= D_r\sigma^2/3$ are the rotational and the translational diffusion coefficients, $\gamma_r = \gamma\frac{\sigma^2}{3}$, $\eta$ is Gaussian white noise variable with $\<\eta\> =0$ and $\<\eta(t)\eta(t')\> = \delta(t-t')$, $F_a$ is the magnitude of the active force acting on the particle, $\B F_{i} = \displaystyle\sum \B f_{ij}$ the forces arising from the interparticle interactions.
In the absence of interaction and noise, particles move with velocity $v_a = F_a/\gamma$, and do not rotate. The control parameters are the volume fraction $\phi$, and the Peclet number ${\Pe \equiv} \frac{v_a}{D_r \sigma} = \frac{v_a\sigma}{3D_t}$, with $\sigma$ the average particle diameter. 
For Brownian spheres, $s = 1$ in Eq.~\ref{eq:newton_r}. 
We develop our theoretical model for arbitrary values of $s$, to allow for a stringent numerical test of our theoretical predictions.

We theoretically determine the spinodal as the limit of stability of a homogeneous system towards the growth of density fluctuations. 
Within the spinodal region, the system is unstable as the diffusivity is $\mathcal D < 0$, so that the flux of particles induced by a concentration gradient,  $J = -{\mathcal D} \nabla \rho$, enhances the gradient in a positive feedback mechanism.
The limit of stability of the homogeneous phase is thus ${\mathcal D} = 0$, or equivalently 
$J = j_{g}-j_{s} = 0$, where $j_g = \rho \tau_{g}^{-1}$ and $j_s = \rho \tau_{s}^{-1}$ are the particle fluxes promoting and suppressing density fluctuations, respectively.

The flux of particles promoting the growth of density fluctuations results from the interparticle collisions inducing a sensible drop in the local pressure. 
In a homogeneous system, only long-lasting interparticle collisions resulting from head-to-head collisions induce such a drop, as we demonstrate in the Appendix~\ref{app:collisions}.
Most of the collisions, therefore, do not promote density fluctuations but rather slow down the particles, endowing them with an effective velocity, $v_e$. Collisions promoting phase separation have, therefore, a frequency $\tg^{-1} \propto \phi v_e$. 
Previous results have demonstrated that, in the homogeneous phase, the effective particle velocity is $v_e = v_a \left(1-\frac{\phi}{\phi^*}\right)$. This density dependence has been related to the pair-correlation function anisotropy~\cite{Zottl2013}, and rationalized in term of the collision rate~\cite{Stenhammar2014}. 
The estimation of the effective velocity allows that of the typical inverse agglomeration time,
\begin{equation}
\tg^{-1} = a \phi \frac{v_e}{\sigma}=a \phi \frac{v_a}{\sigma}\left(1-\frac{\phi}{\phi^*}\right),
\label{eq:tc}
\end{equation}
where $a$ is a constant of order one, and hence to estimate $j_g$. In the limit of stiff particles $\phi^*$ correspond to the close packing volume fraction.

The inverse agglomeration time vanishes for $\phi \to 0$, due to the absence of nearby particles, as well as for $\phi \to \phi^*$.
In this limit particles are stuck, and particles self-propelling in opposite directions are not able to meet and promote a density fluctuation.
We remark that our estimation concerns the agglomeration time in a homogeneous system at volume fraction $\phi$. This time differs from that needed, in a phase-separated state, by a gas-particle to join a cluster~\cite{Redner2013, Buttinoni2013, Wysocki2014, Cates2015, Redner2016}. In particular, in the gas phase far from the critical point the volume fraction dependence of $v_e$ is negligible.

The flux of particles promoting phase separation is contrasted by fluxes promoting the homogeneous phase.
One of these fluxes is driven by the rotational diffusivity of the particles, as two colliding particles might resolve their collision rotating their self-propelling direction. This physical process is the same allowing, in a phase-separated state, particles on the rim of an active cluster to escape from it
~\cite{Redner2013,Buttinoni2013,Wysocki2014,Cates2015}. The inverse timescale of this rotational detaching mechanism is
\begin{equation}
\trd^{-1} = b s D_r,
\label{eq:td1}
\end{equation}
with $b$ a constant of order one. 
In principle, particles may also resolve their collision by diffusing away, giving rise to a flux promoting phase separation driven by the translational diffusivity. However, the phase diagram of ABPs appears~\cite{Fily2012,Fily2014} insensitive to $D_t$, for reasons we rationalize later on. We, therefore, do not consider any stabilizing flux associated with $D_t$.  

The balance of $\tg$ and $\trd$ allows to predict a spinodal line, $\Pe \propto \phi^{-1} (1-\phi/\phi^*)^{-1}$. This prediction capture the numerically observed U-shape of the spinodal line, and the divergence of $\Pe$ in the $\phi \to \phi^*$ limit. However, according to this prediction, the spinodal line also diverges in the $\phi \to 0$ limit; hence, regardless of the volume fraction, a homogeneous system should become unstable and phase separate as $\Pe$ increase. Conversely, previous results indicate that only system with volume fraction larger than a threshold $\phi_m$ becomes unstable.
There is, therefore, an additional stability mechanism in ABPs, which should be relevant at high $\Pe$ and small $\phi$.

\begin{figure}[!!t]
\centering
\includegraphics[width=0.45\textwidth]{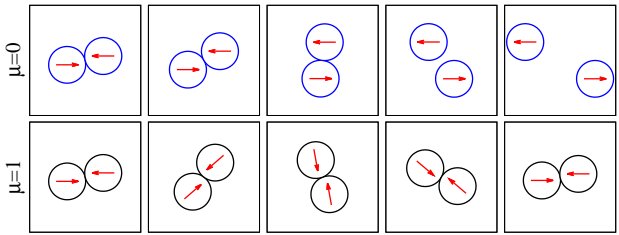}
\caption{
Top row: sliding detaching mechanism of standard frictionless ABPs, as illustrated via the simulation of a collision at high Peclet number. 
Bottom row: the same collision is simulation in frictional ABPs. Friction suppresses the sliding detaching mechanism by inducing the rotation of the self-propelling directions of the colliding particles. Simulations are in the high $\Pe$ limit where stochastic forces are negligible on the considered timescale.
\label{fig:sd}
}
\end{figure}

We identify this mechanisms considering that two particles may resolve their collision without any change in the orientation of their self-propelling direction, but rather by sliding off each other~\cite{Redner2016, Bruss2018}, effectively rotating around their centre of mass. 
This mechanisms is illustrated in the top row of Fig.~\ref{fig:sd}. This physical process is promoted by the activity, which sets the scale of particles' velocity. In a crowded environment, this process is hindered by the density, which slows-down particle motion.
We, therefore, assume particles to slide off each other with a velocity proportional to the effective active velocity.
Thus, the inverse timescale associated with this sliding detaching mechanisms is
\begin{equation}
\tsd^{-1} = c \frac{v_a}{\sigma}\left(1-\frac{\phi}{\phi^*}\right),
\label{eq:td2}
\end{equation}
where $c$ is a constant of order one.

By balancing $j_{g} = \rho \tau_{g}^{-1}$ and $j_{s} = \rho \tau_{s}^{-1} = \rho (\trd^{-1} + \tsd^{-1})$, we determine the spinodal line,
\begin{eqnarray}
\Pe & = & \frac{As}{(\phi^*-\phi)(\phi-\phi_m)}, ~ s \neq 0  \label{eq:model}   \\ 
\phi & = & \phi_m, ~ s = 0  \label{eq:model0}
\end{eqnarray}
with $\phi_m = \frac{c}{a}$ and $A = \frac{b\phi^*}{a}$. Given that $a$, $b$ and $c$ are of order one, so are the $\phi_m$ and $A$, in both 2D and 3D. The critical point is at $\phi_c = \frac{1}{2}(\phi^*+\phi_m)$, $\Pe_c = \frac{4A}{(\phi^*-\phi_m)^2}$. 
The prediction of a vertical spinodal line in the absence of rotational motion, $s=0$, agrees with previous investigations~\cite{Fily2012, Fily2014}. 

\section{Fixing $\phi^*$ and $\phi_m$\label{app:jamming}}
Our theoretical prediction of Eq.~\ref{eq:model} has three free parameters. We have, however, independently estimated both $\phi^*$, the jamming volume fraction, as well as $\phi_m$, the spinodal line in the absence of rotational noise, for the numerical model we consider in the following. 
We find $\phi^*\simeq0.879(0.645)$ in 2D (3D), and $\phi_m\simeq 0.25 (0.345)$, in 2D (3D), as we detail below. Hence, we are left with a theoretical prediction with a single free parameter, the scaling amplitude $A$.

\subsection{Jamming volume fraction, $\phi^*$}
\begin{figure}[!t]
\centering
\includegraphics[width=0.45\textwidth]{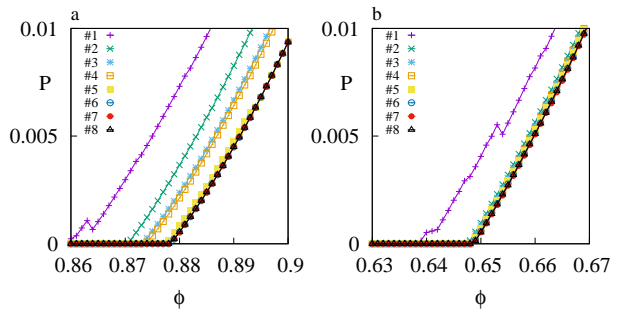}
\caption{Dependence of the pressure on the volume fraction during compression/decompression cycles. For clarity, we only illustrate the compression cycles. Two- and three-dimensional results are illustrated in panel (a) and (b), respectively. 
\label{Fig:jamming}
}
\end{figure}

The jamming volume fraction $\phi^*$ is a protocol-dependent quantity, which activity pushes towards its maximum value. 
To determine $\phi^*$, we cyclically compress and decompress the system across the expected jamming transition, in the absence of motility and noise. 
The volume fraction varies varied in steps of $10^{-3}$, and the energy of the system minimized after every change of volume fraction, via the conjugate gradient protocol. 
Fig.~\ref{Fig:jamming} illustrates subsequent compression curves, in both two and three dimensions. 
The pressure converges after a few cycles to an asymptotic curve, which grows linearly for $\phi > \phi^*$. 
From these results, we estimate $\phi \simeq \phi^* \simeq 0.879$ in two spatial dimensions, and $\phi^* \simeq 0.648$ in three dimensions.

\begin{figure}[!b]
\centering
\includegraphics[width=0.45\textwidth]{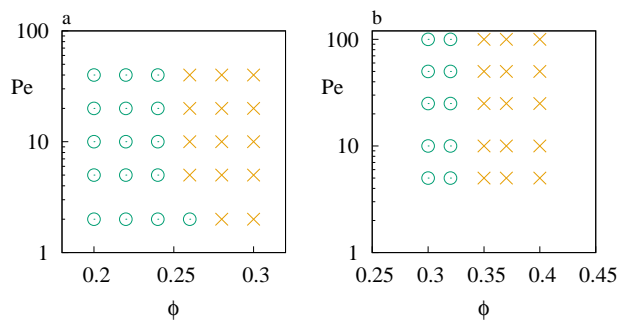}
\caption{Stability phase diagram in the absence of rotational noise. The diagrams have been obtained investigating systems with $N = 32000$ in 2d (left panel) and $N = 64000$ in 3d (right panel). 
\label{Fig:DR0}
}
\end{figure}

\subsection{Phase separation in absence of rotational noise\label{sec:DR0}}
In the absence of rotational noise, $s = 0$, our theoretical model predicts the spinodal line to be Peclet independent, $\phi=\phi_m$. Previous numerical results have investigated this limit, confirming this theoretical prediction~\cite{Fily2012, Fily2014}. 
The investigation of the stability phase diagram for $s = 0$ thus allows estimating $\phi_m$. 

We report our numerical results in Fig.~\ref{Fig:DR0}, for both two and three-dimensional systems. We recovered the Peclet independence of the spinodal line. Deviations from the theoretical predictions occur at small Peclet number, as in this limit thermal diffusivity start being relevant.

From this investigation, we estimate $\phi_m \simeq 0.25$ in two spatial dimensions, and $\phi_m \simeq 0.345$ in three spatial dimensions.

\section{Numerical validation in two- and three spatial dimensions}
\noindent
\begin{figure}[t!]
\centering
\includegraphics[width=0.47\textwidth]{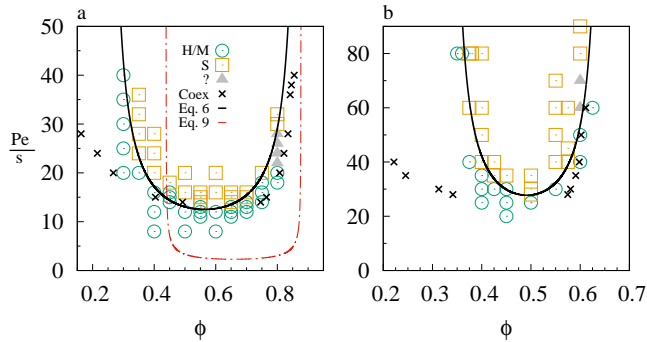}
\caption{
Stability phase diagram for APBs in two (a) and in three (b) spatial dimensions, for $s = 1$.
Circles identify points that either do not phase separate or phase separate via nucleation, within our simulation time. Squares identify points that phase separate via spinodal decomposition.
Triangles identify points for which we were unable to clearly asses the phase.
Stars mark the coexisting volume fractions, as determined from the positions of the peaks of the local density probability distribution.
The fulls lines correspond to the theoretical prediction for the spinodal line of Eq.~\ref{eq:model}. We adjusted the value of $A$ to $1.23$ (2D) and to $0.65$ (3D), and fixed $\phi_m$ and $\phi^*$ as described in the main text.
In panel $(a)$, the dashed line is the theoretical prediction of the continuum approach, Eq.~\ref{eq:continuum}.}
\label{fig:diagram}
\end{figure}
We investigate the phase diagram of ABPs, whose dynamics is governed by Eq.s~\ref{eq:newton_t} and ~\ref{eq:newton_r}. We work in the hard-sphere limit, using the interparticle interaction and the parameter detailed in Appendix~\ref{app:numerics}, and consider systems with up to $N=32000$ particles, in 2D, and up to $N=64000$, in 3D. 
We determine the coexistence line evaluating the position of the peaks of the local density distribution, as discussed in the Appendix~\ref{app:coex}. 
Investigating the dynamics of phase separation~\cite{Redner2013,Stenhammar2014}, and specifically the time dependence of both the fraction of the volume in the low-density phase, and the characteristic size of the density fluctuations, we identify the state points undergoing spinodal decomposition. 
See Appendix~\ref{app:dynamics} for details on the phase separation dynamics at different state points.

Fig.~\ref{fig:diagram}a summarizes our results for standard ($s = 1$) ABPs in two dimensions (2D). In the figure, stars identify the coexistence line, squares points phase separating via spinodal decomposition, and circles points where either nucleation or not phase separation occur.
This phase diagram is consistent with those previously reported in the literature as concern the critical value of the Peclet number, the typical values of the volume fractions, the shape of the coexistence line, and the location of the spinodal region.
Our theoretical prediction of Eq.~\ref{eq:model}, represented as a full black line, correctly delimits the spinodal region. Analogous results for the three dimensional (3D) case are in Fig.~\ref{fig:diagram}b.
\begin{figure}[t!]
\centering
\includegraphics[width=0.47\textwidth]{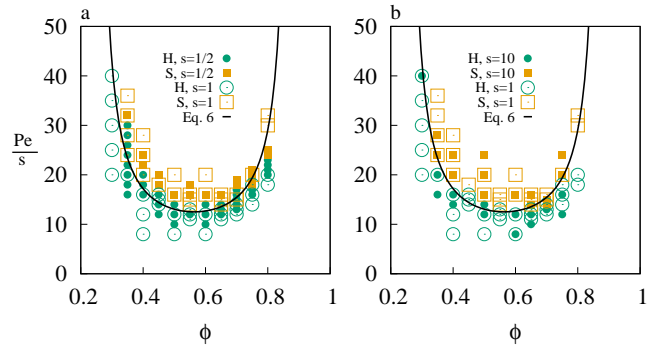}
\caption{
Phase diagram (spinodal lines) of ABPs, in 2D.
With respect to its standard value $s = 1$, the rotational diffusion coefficient is changed by a factor $s = 1/2$ in panel (a), and by a factor $s = 10$ in panel (b).
This does not affect the phase diagram, if the Peclet number is also rescaled by a factor $s$.
}
\label{fig:alpha}
\end{figure}

While these results support our model, they do not clarify whereas our approach, or rather the continuum one, better captures the physics of ABPs. Indeed, a U-shaped spinodal line, which at low density diverges at a finite density, has also been predicted within a continuum description. 
In this approach the coarse-grained density is found to evolve according to a diffusion equation with effective diffusivity~\cite{Fily2012,Fily2014,Bialke2013,Speck2014}
\begin{equation}
\mathcal{D} = D - \frac{\vcg^2(\rho)}{2 s D_r}\left[1 + \frac{d \log \vcg}{d \log \rho} \right],
\end{equation}
where $v_{\rm cg}$ is the coarse-grained velocity along the polarization direction, so that $\mathcal{D} = 0$.
Assuming~\cite{Fily2012,Fily2014,Bialke2013,Speck2014} the coarse-grained velocity to behave as the effective single-particle one, $\vcg(\rho) = v_e(\rho) =  v_a\left(1-\phi/\phi^*\right)$, and interpreting $D$ as density-independent particle diffusivity, this approach predict a spinodal line
\begin{equation}
\Pe_s^{\rm cont} = -s^{1/2} \left[\frac{3}{2} \left(1-\frac{\phi}{\phi^*}\right) \left(1-\frac{2\phi}{\phi^*}\right) \right]^{-1/2},
\label{eq:continuum}
\end{equation}
we illustrate in Fig.~\ref{fig:diagram}a. 
This parameter-free prediction largely underestimates the critical Peclet number, as previously noticed~\cite{Cates2015}.
However, treating $\phi^*$~\cite{Fily2014} (or $\phi_m$~\cite{Speck2014, Speck2016}) as a free parameter, and allowing for a scale factor possibly associated to the density dependence of the particle diffusivity, the prediction of the continuum model becomes comparable to ours, for $s = 1$.

The two theoretical predictions, however, differ as concern the dependence of the spinodal line on the rotational diffusivity, $s D_r$ in our formalism. Indeed, the spinodal line scales linearly with $s$, according to our theoretical prediction of Eq.~\ref{eq:model}, while it scales as $s^{1/2}$ according to the continuum model, Eq.~\ref{eq:continuum}. 
This consideration makes compelling the investigation of the $s$ dependence of the phase diagram, we have performed in 2D. 
We compare the $s=1$ phase diagram with those obtained for $s=1/2$ and $s=10$, in $\Pe/s$--$\phi$ plane, in Fig.~\ref{fig:alpha}. 
In the figure, open symbols correspond to $s = 1$, full ones to $s \neq 1$, and the full black line is as in Fig.~\ref{fig:diagram}a.
The phase diagrams in the $\Pe/s$--$\phi$ are almost indistinguishable. This finding indicates that the spinodal line scales linearly with $s$, and strongly supports our theoretical model.

Furthermore, we notice that the continuum approach predicts a vertical phase boundary in the absence of translational noise~\cite{Fily2014}, regardless of the rotational noise. Conversely, a vertical phase boundary occurs with no rotational noise~\cite{Fily2012, Fily2014}, as predicted by Eq.~\ref{eq:model0}.

\section{Rotational vs translational diffusivity}
Our model, which neglects the role of translational diffusivity, successfully rationalize the motility induced phase diagram. 
We rationalize why the translational diffusivity is irrelevant and the limit of validity of this result, by comparing the diffusion coefficient of the passive suspension to the activity induced effective diffusion coefficients.
For the diffusivity of the passive suspension, we find (2D) $D_p(\phi) = D_t (1-\phi/\phi_d)$ with $\phi_d \simeq \phi^*$, in the volume fraction range we have considered. 
This result is consistent with the expectation for the low-density behavior of Brownian particles~\cite{Dhont1996}. 
We associate two diffusion coefficients to the active suspension, by describing particle motion in the homogeneous phase as resulting from a sequence of steps alternatively taken from distributions corresponding to two different stochastic processes, describing the motion in between collisions and during a collision.  

The stochastic process describing motion in between collisions is that of a persistent random walk, with persistence time $1/sD_r$.
The corresponding diffusivity $D_{\|}$ is evaluated, following previous works~\cite{Fily2012,Stenhammar2014}, considering the steps to have length $l = v_a s^{-1} D_r^{-1}\left(1-\frac{t_c}{t_c+\tmf}\right)$, where $t_c$ and $\tmf$ are the mean duration of a collision, and the mean time between collisions. 
At low density $\tmf \propto (v_a\sigma^{d-1}\rho)^{-1} \gg t_c$, and
$t_c/\tmf = \phi/\phi^*$~\cite{Stenhammar2014}, so that
\begin{equation}
D_{\|} - \frac{D_p(\phi)}{d} \simeq \frac{\sigma^2 D_r \Pe^2}{s} \left(1- \frac{\phi}{\phi^*} \right)^2.
\label{eq:Dp}
\end{equation}
In the above equation, we have taken into account the contribution of the thermal diffusivity, which is divided by a factor $d$ accounting for the fact that $D_{\|}$ is effectively a one-dimensional diffusivity.
The stochastic process describing the motion resulting from the steps performed during the collisions is that of a simple random walk, with step size $\propto \sigma$ and step frequency $1/(t_c+\tmf) \sim 1/\tmf \propto v_a \phi/\sigma$, so that
\begin{equation}
D_{\perp} - \frac{D_p(\phi)}{d}  \simeq \sigma^2 {\Pe} D_r \phi.
\label{eq:Dt}
\end{equation}

\begin{figure}[t!]
\centering
\includegraphics[width=0.47\textwidth]{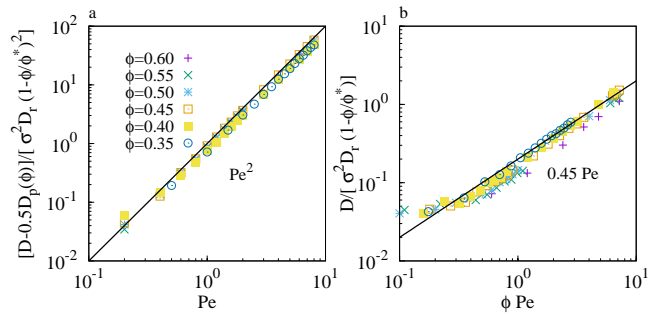}
\caption{
Peclet number dependence of rescaled diffusion coefficients associated to the motion parallel and perpendicular to the self-propelling direction of each particle.
}
\label{fig:diffusion}
\end{figure}

We numerically validate these theoretical predictions by decomposing the instantaneous velocity of particle $i$ in the components parallel and perpendicular to its self-propelling direction, ${\bf v}_i(t) = {\bf v}_i^{\|}(t) + {\bf v}_i^{\perp}(t)$, with ${\bf v}_i^{\|} = (\bf v_i \cdot \bf n_i)\bf n_i$. 
The time integration of these velocities defines a normal and a tangential displacement, ${\bf \Delta r}_{\|,\perp}^i(t) = \int_0^t {\bf v}_i^{\|,\perp}(t) dt$, from which we estimate the diffusion coefficients, $D_{\|,\perp} =  \lim_{t \to \infty }\langle {\bf \Delta r}_{\|,\perp}^2(t)\rangle/2t$.
Fig.~\ref{fig:diffusion} shows that these numerical estimates well compare with the theoretical predictions. 
The theoretical predictions work wells at $\Pe \gtrsim  1$.
Importantly both diffusion coefficients, and in particular $D_{\perp}$ which describes a physical process promoting the homogeneous phase, grow with the Peclet number, and are much larger than the diffusivity of the passive suspension. 
This result explains why the diffusivity of the passive suspension does not influence the motility induced phase diagram.

However, our theoretical prediction fails at small $\Pe$, where the collisional description of the dynamics is no longer appropriate.
Hence, the thermal diffusivity may influence phase separation if the critical point is at $\Pe < 1$, which may occur at very small values of $s$.
In this limit, our theoretical prediction breaks down. 
Indeed, for $s = 0$ our model predicts a vertical phase boundary, and hence a motility-induced phase separation, also in the limit of vanishing motility. With no motility, however, no MIPS occurs and the system behaves as a thermal one~\cite{Digregorio2018}. 
Understanding how the U-shaped spinodal line evolves into a vertical phase boundary as the rotational diffusivity decreases is an interesting avenue of research we leave for the future.

\section{Frictional control of the phase diagram}
The sliding-detaching mechanism occurs when two colliding particles resolve their collision without their self-propelling directions rotating. Interparticle interactions which induce a torque on the particles, therefore, suppress this mechanism, as illustrated in Fig.~\ref{fig:sd}, bottom row. These interactions may result from the shape of the particles, if these are elongated~\cite{Suma2014,Petrelli2018}, and from lubrication forces in wet systems~\cite{Zottl2014}.
In dry systems, frictional forces also induce torques and hence suppress the sliding detaching mechanisms in dry active matter.
Recent results have demonstrated that frictional forces are also present in colloidal hard-spheres suspensions at high enough shear rates, where they give rise to the discontinuous shear thickening phenomenology~\cite{Guy2015, Clavaud2017, Hsu2018, Kawasaki2018}. 
This frictional dependence is rationalized assuming that the hydrodynamic layer, which would give rise to diverging normal forces, break down at a characteristic length set by the particles' surface asperities.
Hence, frictional forces may play a role also in experiments of wet colloidal scale active particles, at high enough Peclet number. 
The tunability of the frictional interaction in colloidal systems~\cite{Hsu2018}, therefore, may  allow to control the motility induced phase separation of these systems.

\begin{figure}[t!]
\centering
\includegraphics[width=0.47\textwidth]{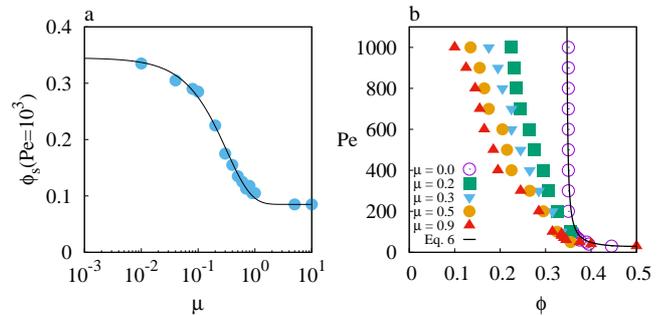}
\caption{
At given $\Pe$, the a homogeneous system becomes unstable towards phase separation at a volume fraction $\phi_s$. 
Panel (a) shows that this volume fraction exponentially approach a limiting value as the friction coefficient increases, $\phi_s(\mu) = \phi_s(\infty) + \Delta\phi_s e^{-\mu/\mu_c}$. For $\Pe = 10^3$ we find $\phi_s(\infty) \simeq 0.085$, $\Delta\phi_s\simeq0.26$ and $\mu_c\simeq0.32$.
Panel (b) illustrate the static friction dependence of the lower spinodal line.
}
\label{fig:mu}
\end{figure}

We investigate the influence of static friction on the motility induced phase separation in three-dimensional numerical simulations, resorting to frictional Mindlin model, as described in the Methods section. 
We find Coulomb's friction coefficient $\mu$ to influence the lower spinodal line, which is critically affected by the sliding detaching mechanisms, not the upper spinodal line.
At each value of the Peclet number, a homogeneous system becomes unstable toward phase segregation at a volume fraction $\phi_s(\Pe,\mu)$ which exponentially decreases with $\mu$, approaching a limiting value, as in Fig.~\ref{fig:mu}a. 
Consistently, the spinodal region widens on increasing the friction coefficient and the Peclet number, as in Fig.~\ref{fig:mu}b.

The combined effect of friction and Peclet number is rationalized considering that the frictional forces, whose strength scales as $\mu v_a \propto \mu \Pe$, can be disrupted by thermal ones, that have a constant magnitude, through an activated process. 
This activated dynamics naturally explains the exponential decay of $\phi_s(\Pe,\mu)$. 
The activation probability decreases as the Peclet number increases, so that the spinodal $\phi_s(\Pe,\mu)$ approaches $\phi = 0$, at all finite values of the friction coefficient, making the frictionless limit a singular one.

These results demonstrate that friction modulates the phase diagram of ABPs, possibly guiding future experiments. Importantly, this modulation occurs as friction suppresses the sliding detaching mechanisms, indirectly confirming the relevance of this mechanism.

\section{Conclusions}
While not representing any experimental system faithfully, active Brownian Particles emerged as a prototypical model exhibiting a motility induced phase separation, and are the standard benchmark for statistical physics of active matter theories.
In ABPs, motility promotes phase separation, as clarified by the MIPS~\cite{Cates2015} or, equivalently, by a mechanistic approach~\cite{Solon2015}: due to the presence of motility, colliding particles are much slower than the others, so that collisions induce a local pressure drop which seeds phase separation. Previous theoretical approaches argued that stochastic processes related to the translational~\cite{Fily2012, Zottl2013, Bialke2013, Speck2014, Fily2014, Stenhammar2014} or rotational~\cite{Buttinoni2013,Redner2013, Redner2016} diffusivity conversely promote the homogeneous phase.

Here we have demonstrated that, besides these previously identified mechanisms, there exists a physical process induced by motility that promotes the homogeneous phase. 
This process dominates over the others at high motility. 
In this limit, the balance of two motility driven processes fixes the spinodal line, which thus becomes motility independent. 
In the opposite limit of small motility, the rotational diffusivity becomes relevant, and the phase diagram becomes motility dependent. 
We have formalized these mechanisms in a kinetic approach and predicted the spinodal line of ABPs up to a scaling amplitude of order one. The estimation of this constant remains an open problem. 

We have explicitly checked that in standard ABPs the translational diffusivity is negligible respect to diffusivities induced by the collisions.
This explain why the translational diffusivity does not affect the phase diagram. 
This scenario, however, certainly changes in the limit of small rotational diffusivity, where thermal and activity induced effects may compete.

The sliding detaching mechanisms we have uncovered involves the coordinated motion of colliding particles. 
Coordinated motion is, by definition, suppressed in active Ornstein–Uhlenbeck~\cite{Klamser2018} and  Monte-Carlo models~\cite{Levis2014}. Accordingly, for these models we expect the spinodal line do diverges in the $\phi \to 0$, as observed. 
Friction also suppresses the sliding detaching mechanism, by inducing the rotation of the self-propelling directions of the particles. 
Our investigation of the effect of friction indicates that this could be used to control the features of the motility induced phase diagram, also in light of recent experimental results~\cite{Hsu2018}.
The frictional dependence is qualitatively rationalized considering that frictional forces scales as $\mu \Pe$. The coexistence of this force scales and of stochastic forces then makes the dynamics of frictional systems an activated one. The quantiative rationalization of the effect of friction on the phase diagram remains, however, an open problem.

\appendix
\section{Numerical details~\label{app:numerics}}
We perform numerical simulations of ABPs in two and three spatial dimensions. 
We use an interparticle interaction model borrowed from the granular community, to model frictional particles; the frictionless model is recovered setting to zero Coulomb's friction coefficient, $\mu$.

The interparticle interaction force has a normal and a tangential component, $\B F_{ij} = \B f_{ij}^{n} +  \B f_{ij}^{t}$.
The normal interaction is a purely repulsive Harmonic interaction, $\B{f}^n_{ij}=k_n(\sigma_{ij}-r_{ij})\Theta(\sigma_{ij}-r_{ij})\B{ \hat r}_{ij}$, $\Theta(x)$ is the Heaviside function, $\sigma_{ij}=(1/2)(\sigma_i+ \sigma_j)$, $\B r_{ij}=\B r_i-\B r_j$, and $\B r_i$ is the position of particle $i$.
The tangential force is $\B{f}^t_{ij}=k_t\vec{\xi}_{ij}$,  where $\B{\xi}_{ij}$ is the shear displacement, defined as the integral of the relative velocity of the interacting particle at the contact point over the duration of the contact, and $k_t=\frac{2}{7}k_n$. 
In addition, the magnitude of tangential force is bounded according to Coulomb's condition: $\left|\B {f}^t_{ij}\right|\leq\mu\left|\B {f}^n_{ij}\right|$.
We work in the hard-sphere limit considering stiff particles, the maximum relative deformation of a particle being $\leq 10^{-4}$ for the range of parameters we have considered. This makes our results insensible to $k_n$, but our numerical investigation more computationally costly than previous ones.

For the frictionless case, $\mu = 0$, and the equation of motion are as in Eq.s~\ref{eq:newton_t},\ref{eq:newton_r}. In the presence of friction, a torque $\frac{ {\B T}_i}{\gamma_r}$ arising from the frictional interparticle interaction is added to Eq.~\ref{eq:newton_r}.

Simulations \cite{Plimpton1995} are done integrating the equation of motion via the overdamped Langevin algorithm, with integration timestep $2\times{10}^{-8} D_r^{-1}$.\\

\section{
Agglomeration timescale and effective particle velocity \label{app:collisions}}
Our kinetic model requires the estimation of the agglomeration timescale $\tg$, which is the average time a particle waits before experiencing a collision promoting the formation of a cluster. 
This time scale depends on the typical particle velocity. We have assumed this to be an effective density-dependent velocity, rather than the bare particle velocity $v_0$. Here we provide data supporting this assumption.

If all collisions promote the formation of a cluster, then the timescale of interest is the mean free time which, in the ABP context, depends on the density and the bare particle velocity $v_0$. In ABPs, however, only the long-lasting collisions able to sensibly slow-down particle motion, hence reducing the local pressure, and possibly seeding the formation of a cluster.

\begin{figure}[!t]
\centering
\includegraphics[width=0.48 \textwidth]{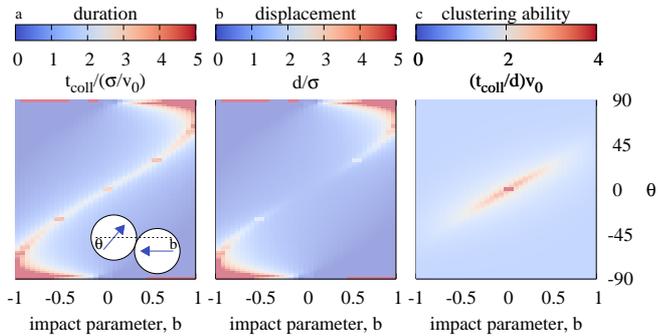}
\caption{Panel (a) and (b) illustrate the dependence of the duration of a collision, $\t$, and of the average of the displacements of the colliding particles, $d$, on the impact parameter, $b$, and the relative angle between the self-propelling direction, $\theta$. These quantities are defined in the inset of panel $a$, and $\theta = 0$ corresponds to a head-to-head collision. Panel (c) illustrates the $b$ and $\theta$ dependence of the clustering ability $\t/d$.
Quantities are non-dimensionalised using the bare particle velocity $v_0$ and the particle diameter $\sigma$.
\label{Fig:collision}
}
\end{figure}

To evaluate how may collisions could potentially lead to the formation of a cluster, we have investigated the dynamics of two-particle collisions, as a function of the impact parameter $b$ and of the relative angle of the self-propelling directions of the two particles, $\theta$. 
The inset of Fig.~\ref{Fig:collision}a defines these quantities. In these simulations, there is no rotational noise.

Figure~\ref{Fig:collision} illustrates in panel (a) the b and $\theta$ dependence of the collision duration, $\t$.
The line of maximal values correspond to $\theta(b) = \arctan\left(\frac{b}{\sqrt{1-b^2}}\right)$, but long lasting collisions also occur for $|\theta| \simeq \pm 90$ and $|\theta| \simeq \pm 180$ (not shown). 
Panel (b) illustrates the average displacement of the particles during the collision, $d = \frac{1}{2}\sum_{i=1}^2 \left[(x_i(\t)-x_i(0))^2+(y_i(\t)-y_i(0))^2\right]^{1/2}$.

Collisions able to sensibly slow down particle motion are those with large $\t$ and small d. Hence, we consider the ratio $\t/d$ as a proxy of how likely a collision acts as a cluster seed. 
Panel (c) shows that this clustering ability is strongly peaked around collisions with $\theta(b) = \arctan\left(\frac{b}{\sqrt{1-b^2}}\right)$, and $b$ small.

Since the clustering ability sharply peaks around some characteristic values of the impact parameters, only a tiny fraction of all collisions induce the formation of a cluster. Because of this, in between two collisions promoting the formation of a cluster, a particle experiences many other collisions. These collisions slow down the particle motion endowing the particles with an effective velocity. We argue that this effective velocity, rather than the bare one, sets the agglomeration time scale.

\section{Determination of the spinodal region\label{sec:cg}}
The spinodal line is a mean-field concept, and in finite systems, the separation between nucleation and spinodal decomposition is not sharp. Nevertheless, on increasing the system size, the crossover between the two different phase separation mechanisms allows for a meaningful operative identification of the state points where phase separation occurs via spinodal decomposition.

To identify the state points within the spinodal region, we have first identified those that phase separate after relatively short simulations, investigating the distribution of the coarse-grained density, as described in Appendix~\ref{sec:cg} below.
Then, for a relevant subset of those points, we have investigated the dynamics of phase separation, to distinguish between nucleation and spinodal decomposition, as described in Appendix~\ref{app:dynamics}.

\subsection{Coarse-grained density~\label{app:coex}}
\begin{figure}[!b]
  \begin{center}
   \includegraphics[width=0.4\textwidth]{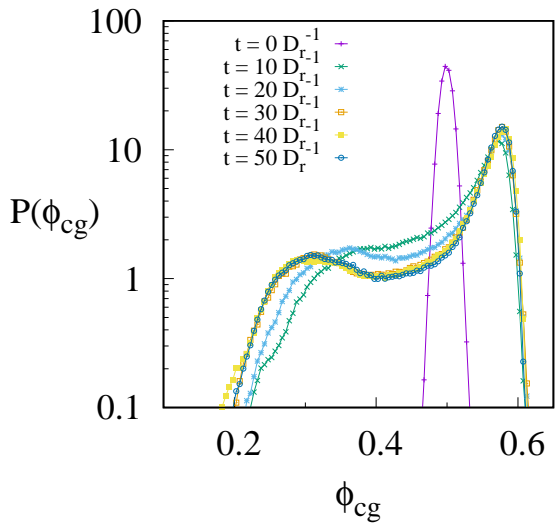}
  \end{center}
  \caption{Time evolution of the local volume fraction distribution, in three dimensions, at $\phi = 0.5$ and $\Pe = 30$, for $N = 32000$.
\label{Fig:EvolPvf}}
\end{figure}
We determine if a system is homogeneous or phase separated investigating the probability distribution of the coarse-grained density, $\rho_{cg}(r)$, or equivalently of the coarse-grained volume fraction,  $\phi_{cg}(r) = \rho_{cg}(r) \langle v \rangle$, with 
$\langle v \rangle$ average particle volume.
Following Ref.~\cite{Cates2013}, we define $\rho_{cg}(r)$ by convoluting the number density $\sum_i\delta({\bf r}-{\bf r_i})$ with $f({\bf r})=Z\exp[-1/(1-r^2/w^2)]$, with $w = 3.5\sigma$ and $Z$ a normalization factor.
Figure~\ref{Fig:EvolPvf} illustrates that in simulations started from a homogeneous configuration, $\rho_{cg}(r)$ evolves until it converges to a steady-state distribution, which in the figure is bimodal. In the range of parameters we have considered, convergence occurs within $100 D_r^{-1}$.

\begin{figure}[t!]
\centering
\includegraphics[width=0.47\textwidth]{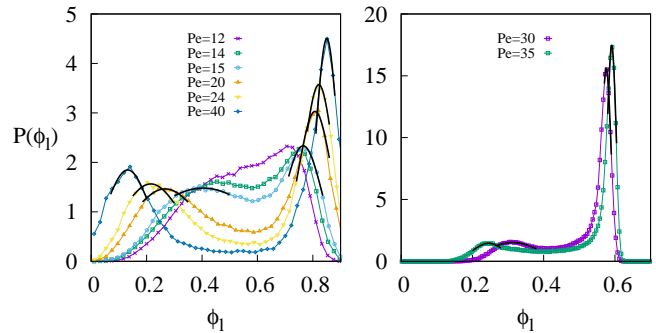}
\caption{Peclet number dependence of the probability distribution of the local volume fraction $\phi_{\rm cg}$, in two (left) and three dimensions (right). Full black lines are local fits to Gaussian function used to estimate the coexisting densities.
In two dimensions, $N = 32000$, while in three dimensions $N=64000$.
\label{Fig:Pvf}
}
\end{figure}

We consider a point in the $\phi-\Pe$ to be phase-separated if the probability distribution of the local density is bimodal, and if doubling the system size yields consistent results.
In two dimensions, we consider systems with $N$ up to $32000$, in three dimensions with $N$ up to $64000$. 
Fig.~\ref{Fig:Pvf} illustrates example distributions.

We evaluate the coexisting densities by locally approximating the distribution via Gaussian functions. The results of these fits are illustrated in the figure.
The coexisting volume fractions are in Fig. 1 of the main text.

Note that, while the two panels of Fig.~\ref{Fig:Pvf} refer to volume fractions which are close to the critical ones, the height of the two peaks is sensibly different. This difference occurs as the coexistence curve is extraordinarily flat and asymmetrical close to the critical point, in particular in three dimensions. 

\subsection{Dynamics of phase separation\label{app:dynamics}}
\begin{figure*}[!t]
\begin{center}
\includegraphics[width=0.65\textwidth]{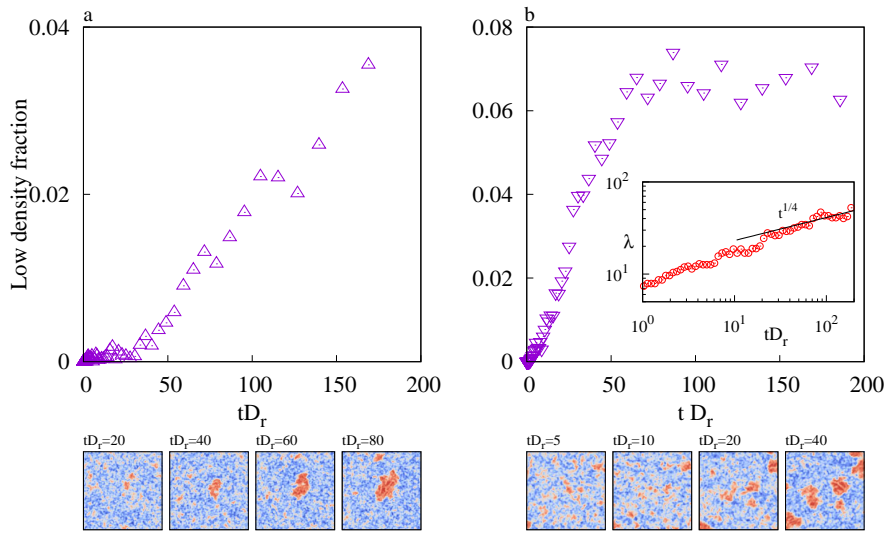}
\end{center}
\caption{Evolution of the fraction of the overall volume with a density smaller than the low-density coexistence density, for a $N=32000$ particle system, in two dimensions. 
In panel (a) $\phi = 0.35$, while in panel (b), $\phi = 0.4$. In both panels, $\Pe = 20$.
At $\phi = 0.35$ (a) the low density fraction only increases after a transient. This indicates that phase separation occurs via nucleation. At $\phi = 0.4$ (b) the fraction grows quickly at early times, indicating that the phase separation proceeds via spinodal decomposition.
Afterword, the system coarsens. 
The growth of the length scale of the density fluctuations is compatible with the expected $\lambda \sim t^{1/4}$ law, as illustrated in the inset. 
The bottom panels illustrate maps of the coarse grained density distribution, which consistently suggests that phase separation occur via different processes at the different volume fractions. Note the different timescales.
\label{Fig:gf}
}
\end{figure*}
\begin{figure*}[!t]
\begin{center}
\includegraphics[width=0.55\textwidth]{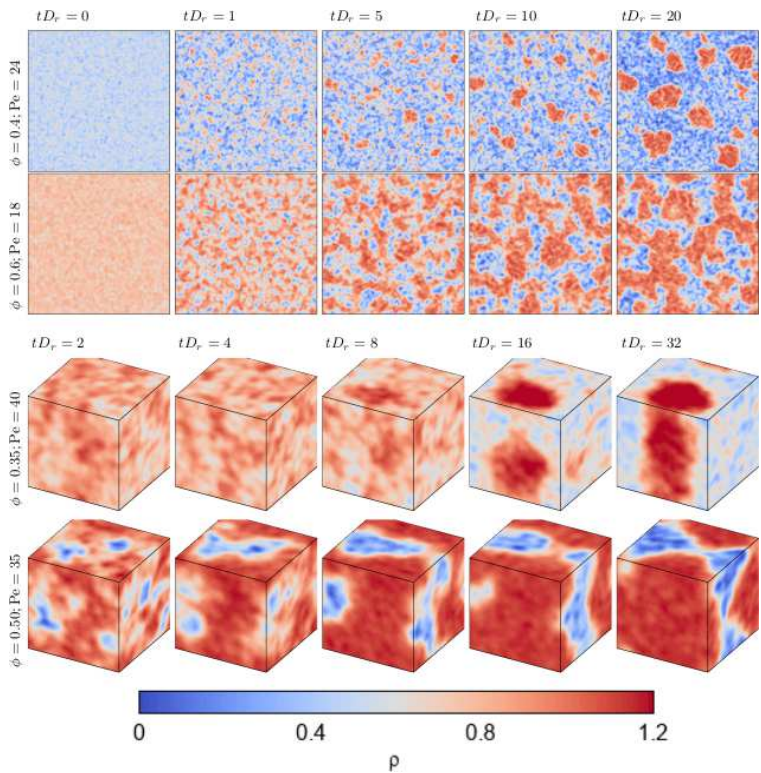}
\end{center}
\caption{Time evolution of the coarse-grained density, for systems of $N = 32000$ (2d) and $N=64000$ (3d) particles. The three rows refer to different state points in the $(\phi,{\rm Pe})$ plane, just above what we have identified as the spinodal line.
We consistently observe the spinodal decomposition of the system.
\label{Fig:spinodal}
}
\end{figure*}
We investigate the dynamics of phase separation to rationalize whether a system undergoes segregate via spinodal nucleation rather than via nucleation.
First, we consider the time dependence of the percentage of the total volume with local volume fraction smaller than that of the coexisting low-density phase, $\alpha(t) = V[{\phi_{\rm cg} \leq \phi_{\rm coex}}(t)]/V_{\rm tot}$. 
If phase separation proceeds via spinodal decomposition, then $\alpha(t)$ quickly varies at short times, being the homogeneous system unstable. Conversely, if phase separation occurs via nucleation, then $\alpha(t)$ only starts varying after a transient, corresponding to the nucleation time. 

Fig.~\ref{Fig:gf} illustrates the result of this investigation, in two dimensions. 
Panel (a) and (b) refer to different state points, that have the same $\Pe$ and differ in volume fraction by $\Delta \phi = 0.05$.
At $\phi = 0.35$, phase separation is seen to occur via nucleation, while at $\phi = 0.4$, it occurs via spinodal decomposition. The associated snapshots of the local density field confirm this interpretation. 

Then,  we investigate the time evolution of the typical length-scale $\lambda$ of the density fluctuations. We define $\lambda$ as the distance at which the correlation function of the coarse-grained density first becomes zero. The inset of panel (b) shows that $\lambda$ grows as a power-law with time at $\phi = 0.4$ and $\Pe = 20$. The growth exponent is compatible with $1/4$, the expected exponent for the coarsening exponent in two-dimensional systems with conserved order parameter. This result further supports our interpretation, namely that at $\phi = 0.4$ and $\Pe = 20$ the system phase separates via spinodal decomposition.

We provide more examples of systems phase separating via spinodal decomposition, for state points close to our identified spinodal line, in both two and three dimensions, in Fig.~\ref{Fig:spinodal}.

\begin{acknowledgments}
P.N., J.C. and MPC acknowledges support from the Singapore Ministry of Education through the Academic Research Fund (Tier 2) MOE2017-T2-1-066 (S) and from the National Research Foundation Singapore, and are grateful to the National Supercomputing Centre (NSCC) of Singapore for providing computational resources.
\end{acknowledgments}

\newpage

\end{document}